\documentclass[twocolumn,showpacs,prl]{revtex4}% Physical Review Letters

\usepackage{graphicx}% Include figure files
\usepackage{dcolumn}% Align table columns on decimal point
\usepackage{bm}% bold math
\usepackage{color}

\begin{document}

\preprint{???}

\title{Dispersion relation of lipid membrane shape fluctuations by neutron spin-echo spectrometry}

\author{Maikel~C.~Rheinst\"adter$^1$}\email{rheinstaedter@ill.fr}
\author{Wolfgang~H\"au{\ss}ler$^2$}
\author{Tim~Salditt$^3$}

\affiliation{$^1$Institut Laue-Langevin, 6 rue Jules Horowitz, BP
156, 38042 Grenoble Cedex 9, France\\
$^2$FRM-II, Technische
Universit\"at M\"unchen, Lichtenbergstra{\ss}e 1, 85747 Garching,
Germany\\
$^3$Institut f\"{u}r R\"{o}ntgenphysik, Friedrich-Hund Platz 1,
37077 G\"{o}ttingen, Germany}

\date{\today}% It is always \today, today,
             %  but any date may be explicitly specified

%------------------------------------------------------------------------------------
\begin{abstract}
We have studied the mesoscopic shape fluctuations in aligned
multilamellar stacks of DMPC bilayers using the neutron spin-echo
technique. The corresponding in plane dispersion relation
$\tau^{-1}$(q$_{||}$) at different temperatures in the gel
(ripple, P$_{\beta'}$) and the fluid (L$_{\alpha}$) phase of this
model system has been determined. Two relaxation processes, one at
about 10ns and a second, slower process at about 100ns can be
quantified. The dispersion relation in the fluid phase is fitted
to a smectic hydrodynamic theory, with a correction for finite
q$_z$ resolution. We extract values for, the bilayer bending
rigidity $\kappa$, the compressional modulus of the stacks $B$,
and the effective sliding viscosity
$\eta_3$. %This technique offers a novel approach to determine the
%elasticity parameters in phospholipid bilayers by direct
%measurement of dynamical properties.
The softening of a mode which can be associated with the formation
of the ripple structure is observed close to the main phase
transition.
\end{abstract}

% 87.16.Dg Membranes, bilayers, and vesicles
% 83.85.Hf X-ray and neutron scattering
% 62.20.Dc Elasticity, elastic constants

\maketitle

%------------------------------------------------------------------------------------
%Introduction
%----------------------------------------------------------------------
\begin{figure*}
\centering
%\resizebox{1.00\columnwidth}{!}{\rotatebox{0}{\includegraphics{diffuse_sheet_mit_rocking.eps}}}
\resizebox{0.70\columnwidth}{!}{\rotatebox{0}{\includegraphics{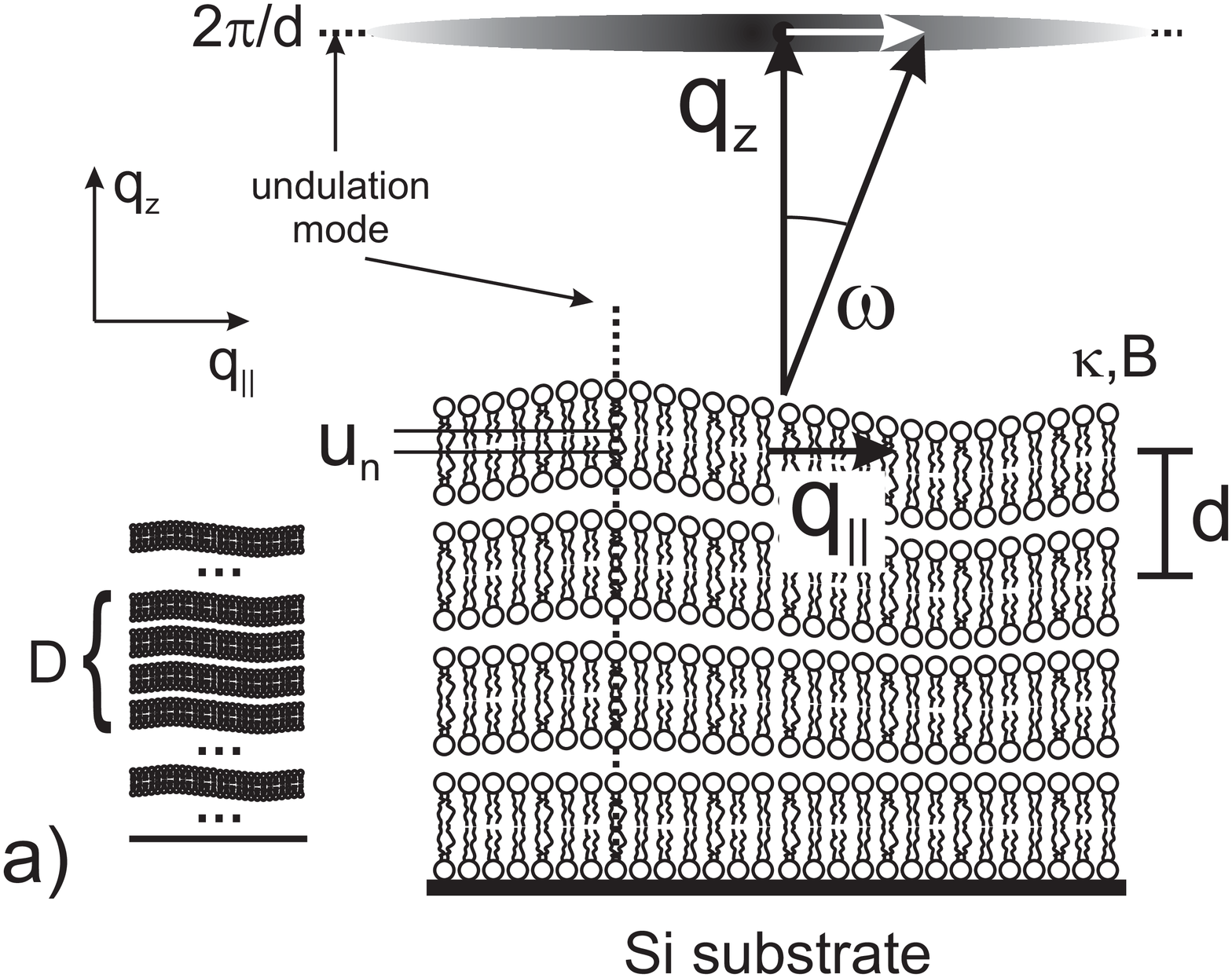}}}
\resizebox{0.75\columnwidth}{!}{\rotatebox{0}{\includegraphics{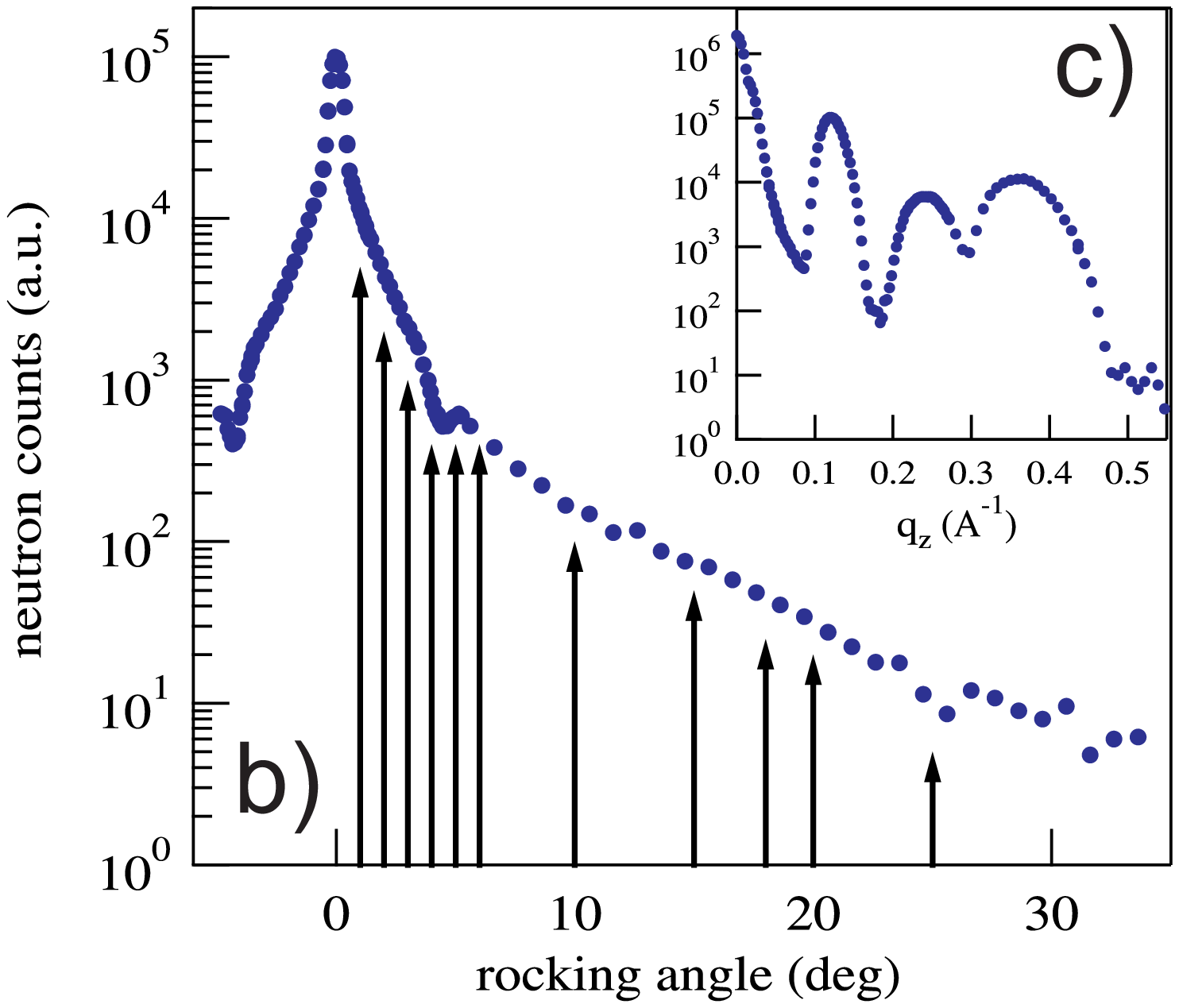}}}
\begin{minipage}[b]{5.1cm}
\centering \caption[]{(Color online). (a) Sketch of the scattering
geometry. (b) Rocking scan at T=30$^{\circ}$C through the first
Bragg peaks and along the diffuse Bragg sheet in log-log scaling.
The arrows exemplary mark positions of inelastic scans in the
fluid phase. (c) Inset shows a reflectivity curve as measured on
IN11 with relaxed momentum resolution
($\Delta\lambda/\lambda\simeq$15\%).}
\label{rocking_fluid_graph.eps}\end{minipage}
\end{figure*}
\begin{figure*}
\centering
\resizebox{0.31\textwidth}{!}{\rotatebox{0}{\includegraphics{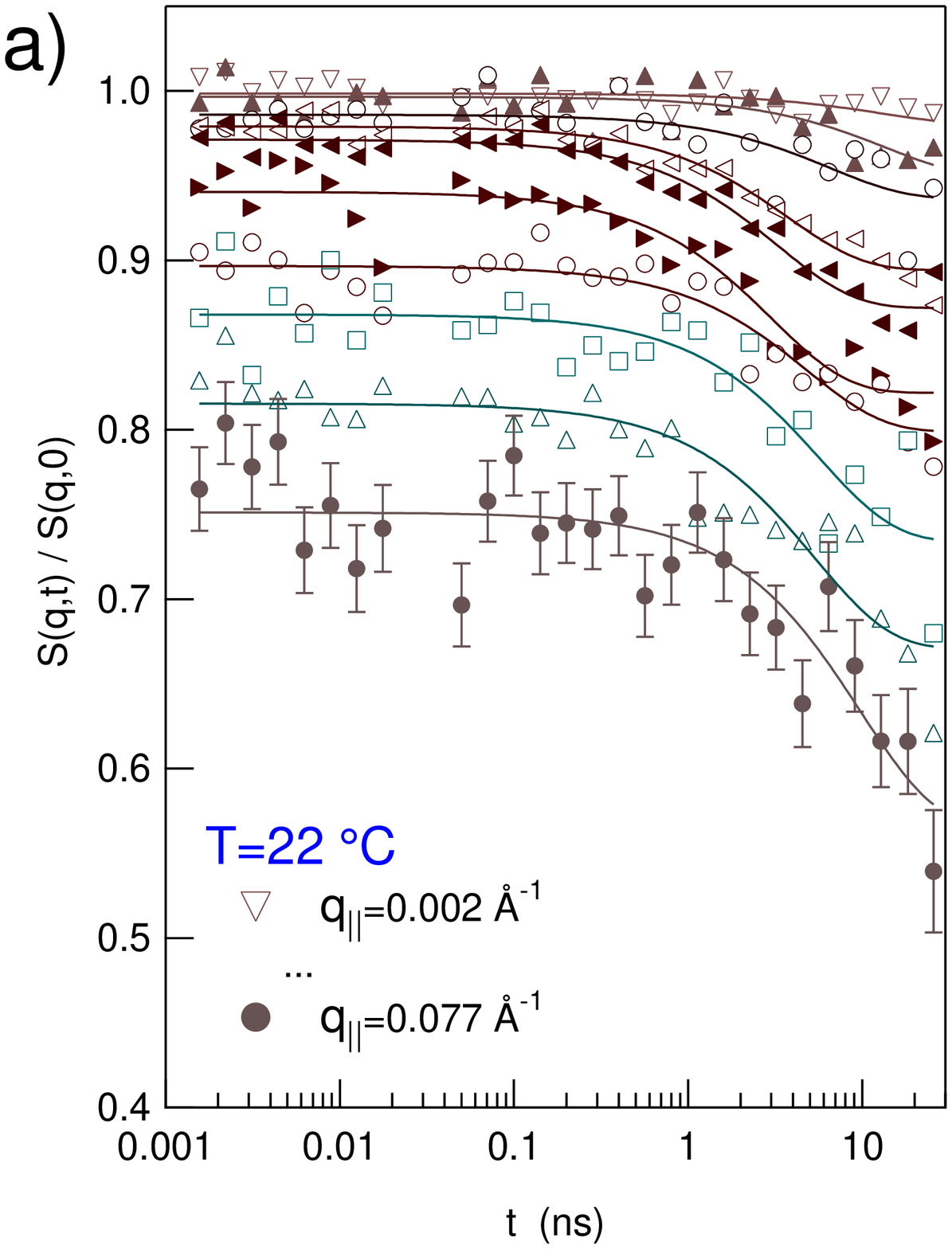}}}
\resizebox{0.31\textwidth}{!}{\rotatebox{0}{\includegraphics{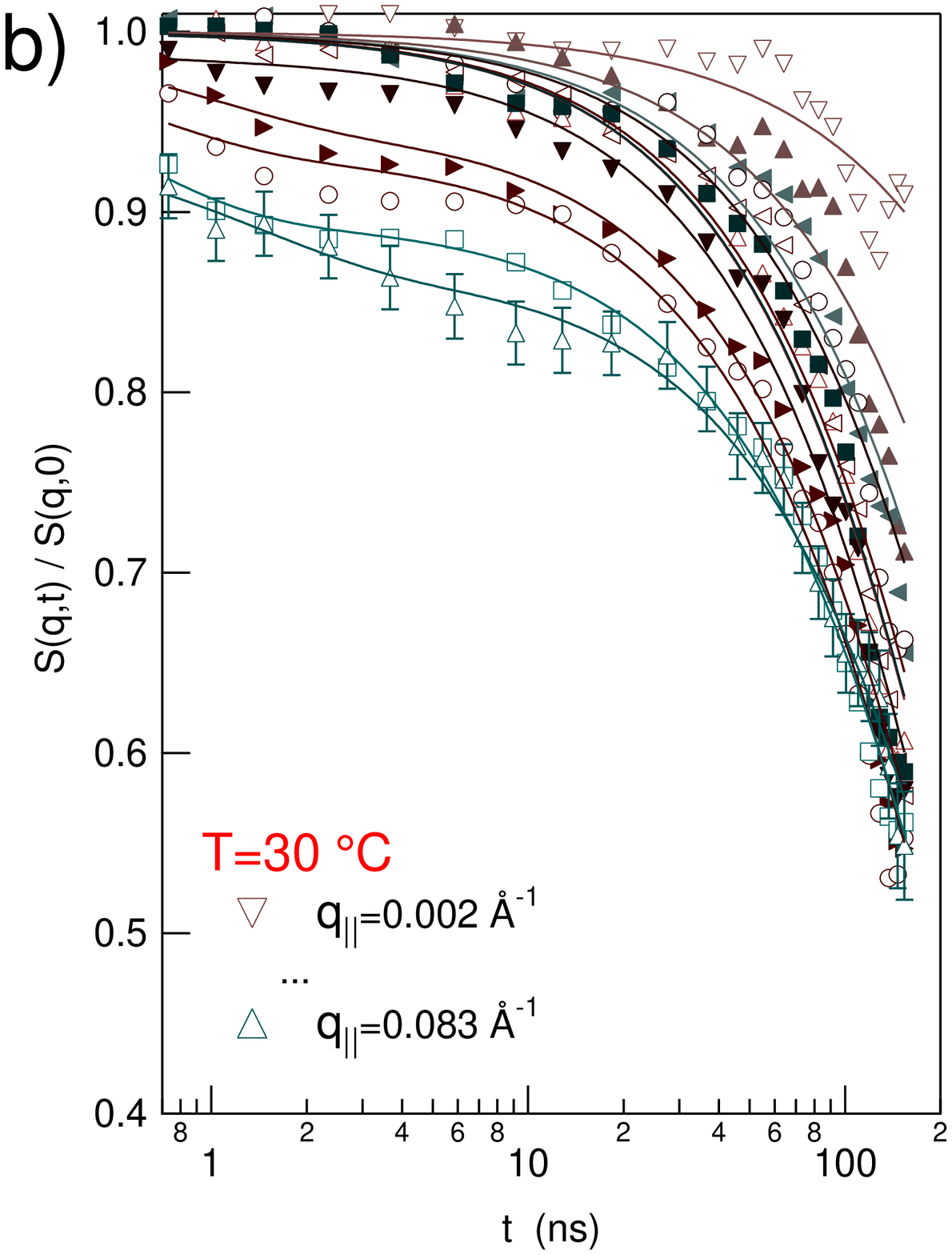}}}
\resizebox{0.31\textwidth}{!}{\rotatebox{0}{\includegraphics{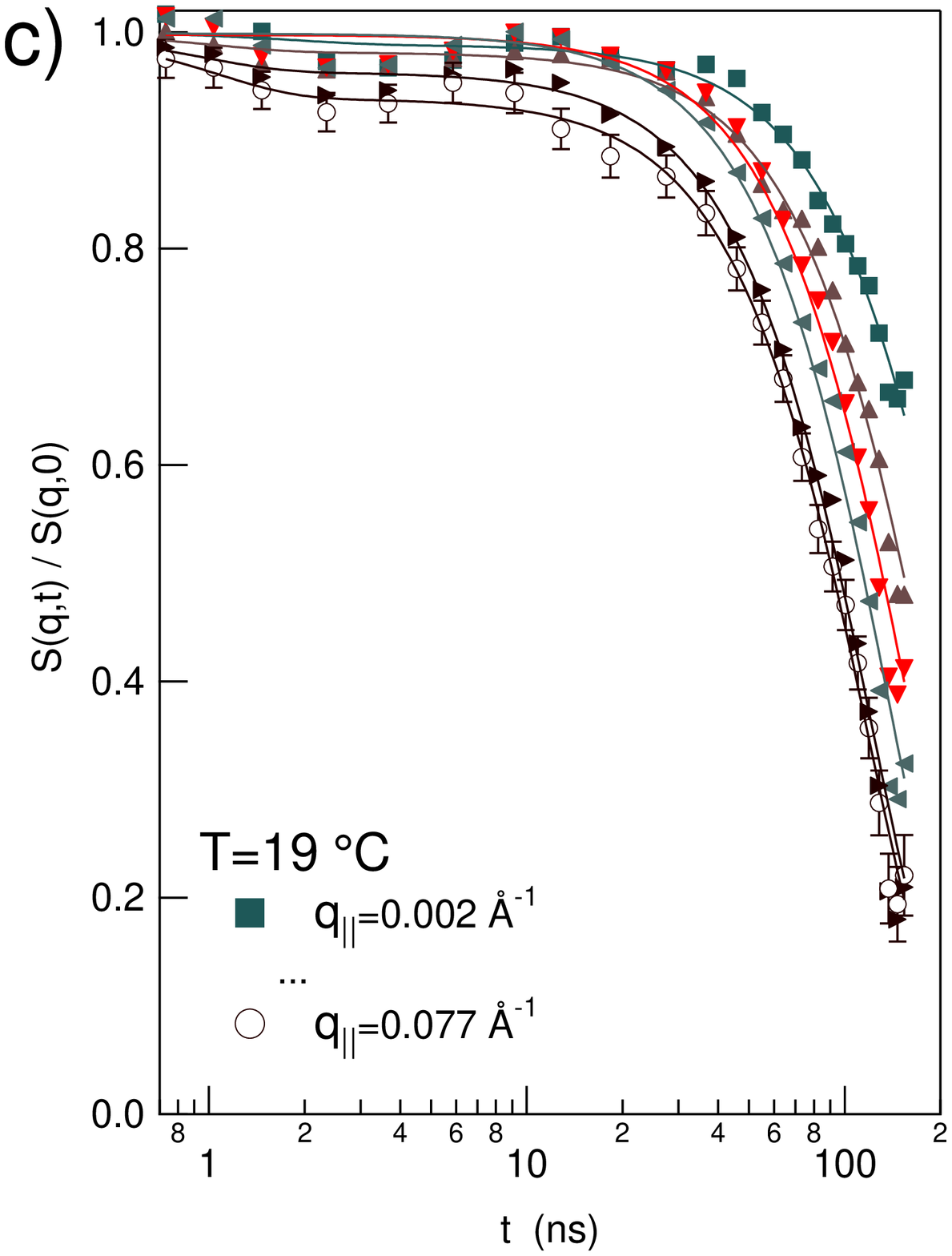}}}
\caption[]{(Color online). Intermediate scattering function
S(q,t)/S(q,0) for selected q$_{||}$ values in the interval
0.002\AA$^{-1}<$q$_{||}<$0.8\AA$^{-1}$ measured on (a) IN11 at
T=22$^{\circ}$C, just above the temperature of the main phase
transition in deuterated DMPC -d54, (b) on IN15 at 30$^{\circ}$C,
in the fluid phase of the phospholipid bilayers, and (c) on IN15,
at 19$^{\circ}$C in the gel phase. Two relaxation processes, one
at about 10ns and a second slower process at about 100ns are
clearly distinguished. Solid lines in the Figures represent
least-square fits to two exponential
decays.}\label{membrane_fluid_plot.eps}
\end{figure*}

Thermally excited shape fluctuations of membranes are readily
observed in the light microscope, e.g., the flickering of lipid
vesicles or biological membranes. On the nanometer scale,
undulations of the lipid bilayer are experimentally much more
difficult to probe. Few spectroscopic techniques offer the
 spatial and temporal resolution needed for the validation of
the theoretic work on collective bilayer motion, see e.g.,
\cite{Seifert:1993,Romanov:2002,Zilman:1996,Lipowsky:1995,Lindahl:2000}.
Up to now, thermal fluctuations of lipid membranes, in particular
in multilamellar stacks, have been mainly investigated by thermal
diffuse scattering, i.e., x-ray lineshape analysis
\cite{Safinya:1986,Petrache:1998}. Elastic scattering has led to a
detailed understanding of the static properties of thermal
fluctuations in lipid membranes and the elasticity properties
governing these fluctuations. According to linear smectic
elasticity theory \cite{Caille:1972,Lei:1995} thermal fluctuations
in the fluid phase of the membrane are governed by the free energy
functional
%(Hamiltonian) \cite{Caille:1972,Lei:1995,LeiThesis:1993}
(Hamiltonian)
\begin{equation}
H = \int_A d^2r
\sum_{n=1}^{N-1}\left(\frac{1}{2}\frac{B}{d}(u_{n+1}-u_n)^2+\frac{1}{2}\kappa\left(\nabla^2_{||}u_n\right)^2\right)~,
\label{Hamiltonian}\end{equation} where $\kappa$ denotes the
bilayer bending rigidity, $A$ the area in the $xy$-plane, $N$ the
number of bilayers, and $u_n$ the deviation from the average
position $n~d$ of the $n$-th bilayer, $d$ is the lamellar spacing.
$B$ and $K = \kappa/d$ are elastic coefficients, governing the
compressional and bending modes of the smectic phase,
respectively. A fundamental length scale in these systems is given
by the smectic penetration length $\Lambda = \sqrt{K/B}$. Aligned
lipid bilayers allow a separate determination of both parameters
$K$ and $B$ \cite{Lyatskaya:2001,Salditt:2003}. However, a full
understanding of the collective dynamics should include
experimental determination of relaxation rates, the relevant
transport coefficients (viscosities), and in particular the
characteristic dispersion relation of the relevant modes.

In this letter we present an experimental dispersion relation for
aligned multilamellar lipid membranes $\tau^{-1}(q_{||})$ as a
function of lateral momentum transfer $q_{||}$ (in the plane of
the bilayers), measured by neutron spin-echo spectrometry (NSE).
From analysis of the dispersion relation, the effective sliding
viscosity, $\eta_3$, as well as the static properties $\kappa$ and
$\Lambda$ are obtained. A softening of a mode which can be
associated with the formation of the ripple structure is observed
close to the main phase transition.  In the $q_{||}$ range probed,
the bilayer displacement $u_n$ can be assumed to be a continuous
variable. It is therefore not sensitive to the discrete molecular
structure, in contrast to the dynamics on the molecular length
scale (acyl chain distance) which can be probed by inelastic
neutron scattering using three-axes spectrometry
\cite{RheinstaedterPRL:2004}. We have selected NSE for this study
since the undulation modes at high $q_{||}$ are too fast to be
accessed by x-ray photon correlation spectroscopy (XPCS) and the
lateral length scales are to small to be resolved by dynamic light
%scattering (DLS) \cite{Hirn:1999,Hildenbrand:2005}. XPCS has
scattering (DLS) \cite{Hildenbrand:2005}. XPCS has successfully
been used for investigations of free-standing thermotropic liquid
crystals, which exhibit much slower acoustic modes, corresponding
to center of mass movement of the entire film
\cite{SikharulidzePRL:2003}. NSE has been
%\cite{SikharulidzePRL:2003,SikharulidzePRE:2005}. NSE has been
used before to study phospholipid membranes,
%\cite{Pfeiffer:1989,Pfeiffer:1993,Takeda:1999}. However, a
\cite{Pfeiffer:1989,Takeda:1999}. However, a dispersion relation
over a wide range in $q_{||}$ in the fluid multilamellar state has
not been achieved so far, in part due to flux and technical
limitations in the past.

%----------------------------------------------------------------------
%Method
%----------------------------------------------------------------------
NSE directly measures the intermediate scattering function S({\bf
q},t), which is related to the density distribution $\rho({\bf
r},t)$ in the sample by
\begin{equation}
S({\bf q},t)=\int d^3{\bf R} e^{-i{\bf qR}}\int d^3{\bf
r}\left\langle\rho({\bf r},0)\rho({\bf r+R},t)\right\rangle.
\label{scatteringfunction}\end{equation} %The corresponding decay
%of polarization in the NSE data is usually fitted to a stretched
%exponential $\exp{[-(t/\tau)^{\beta}]}$, where $\beta$ is the
%stretching exponent.
The spin-echo experiments were carried out at the IN11 and IN15
spectrometers, situated at the cold source of the high flux
reactor of the Institut Laue-Langevin (ILL) in Grenoble, France.
Wavelength bands centered at $\lambda$=7.4\AA\ and
$\lambda$=14\AA\ with $\Delta \lambda/ \lambda \simeq 0.15$
(FWHM), respectively, have been set by a velocity selector.
%To optimize the flux on the sample the first slit behind the
%supermirror polarizer was opened to 30 mm (horz) x 30 mm (vert),
%accepting the full beam of the guide.
%The beam size at the sample was reduced by an Cd aperture to 10 mm
%(horz) x 40 mm (vert), matched to the geometry of the sandwich
%sample.
%----------------------------------------------------------------------
%sample
%----------------------------------------------------------------------
%Multilamellar samples were composed of  several thousands of lipid
%bilayers of DMPC (1,2-dimyristoyl-sn-glycero-3-phoshatidylcholine)
%with deuterated chains, aligned on silicon substrates, and
%separated by layers of water (D$_2$O), resulting in a structure of
%smectic A symmetry.
Partially (acyl chain) deuterated DMPC-d54
(1,2-dimyristoyl-sn-glycero-3-phoshatidylcholine) was obtained
from Avanti Polar Lipids. Highly oriented multi lamellar membrane
stacks of several thousands of lipid bilayers were prepared by
spreading lipid solution of typically 25mg/ml lipid in
trifluoroethylene/chloroform (1:1) on 2'' silicon wafers, followed
by subsequent drying in vacuum and hydration from D$_2$O vapor
\cite{Muenster:1999}, resulting in a structure of smectic A
symmetry. Twenty such wafers separated by small air gaps were
combined and aligned with respect to each other to create a
''sandwich sample'' consisting of several thousands of highly
oriented lipid bilayers (total mosaicity about 0.5$^{\circ}$),
with a total mass of about 400mg of deuterated DMPC. During the
experiment, the samples were kept in a closed temperature and
humidity controlled Aluminum chamber, and were hydrated from the
vapor phase.

%----------------------------------------------------------------------
%experimental results
%----------------------------------------------------------------------
The layer fluctuations of lowest energy are the undulation modes,
i.e., highly correlated (conformal) layer displacements which keep
the inter-layer distances approximately constant. This
conformality explains the strong modulation of the diffuse
scattering in $q_z$, i.e., the Bragg sheet structure
\cite{Salditt:2003}, corresponding to the diffuse tails (at
$q_{||}
>0$) of the specular Bragg peaks  ($q_z=n ~2\pi/d, q_{||}
=0$). We have now measured spin-echo curves on the first Bragg
sheet as a function of $q_{||}$, i.e., at constant $q_z=2\pi/d$. A
sketch of the scattering geometry is shown in
Fig.~\ref{rocking_fluid_graph.eps} (a). The in plane component of
the scattering vector is calculated to
q$_{||}$=q$_{z}\tan(\omega)$, with q$_{z}\approx 0.11$\AA$^{-1}$
for the first reflectivity Bragg peak. $\omega$ is the rocking
angle, i.e., the sample rotation with respect to specular Bragg
angle. Figure~\ref{rocking_fluid_graph.eps} (b) depicts a rocking
curve in the fluid phase at T=30$^{\circ}$C on the first
reflectivity Bragg peak where the sharp Bragg component and the
diffuse Bragg sheet can be well distinguished. In the inset, (c),
a reflectivity curve is plotted. Peaks are distinctly broadened by
the wavelength distribution.

\begin{figure}
\centering
\resizebox{0.90\columnwidth}{!}{\rotatebox{0}{\includegraphics{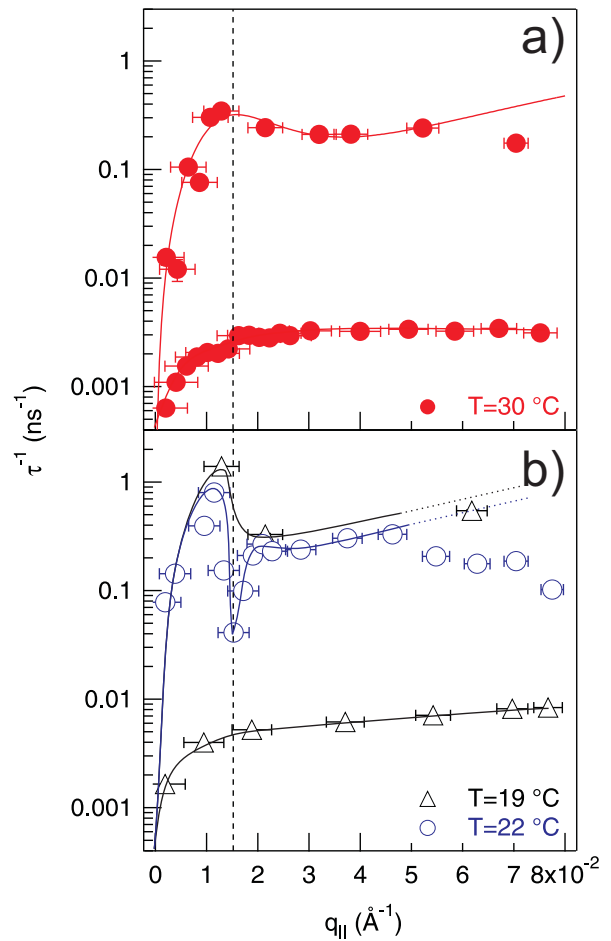}}}
%\resizebox{1.00\columnwidth}{!}{\rotatebox{0}{\includegraphics{dispfit_ribotta_cdr.eps}}}
%\resizebox{1.00\columnwidth}{!}{\rotatebox{0}{\includegraphics{dispersion_fluid_graph_cdr.eps}}}
%\resizebox{1.00\columnwidth}{!}{\rotatebox{0}{\includegraphics{dispfit_ribotta_ripple2_cdr.eps}}}
\caption[]{(Color online). (a) Dispersion relations at
T=30$^{\circ}$C. The solid line is a fit to Eq.~(\ref{ribotta}).
(b) Dispersion relations in the gel (19$^{\circ}$C) and in the
fluid phase (22$^{\circ}$C). A pronounced soft mode is observed at
q$_0\approx$0.015\AA$^{-1}$ at 22$^{\circ}$C (dotted vertical
line). Solid lines in (b) are guides to the eye.}
\label{dispersion_fluid_graph.eps}
\end{figure}

Figure~\ref{membrane_fluid_plot.eps} shows the intermediate
scattering function S(q$_{||}$,t) for selected q$_{||}$ values for
spin-echo times of 0.001ns$<$t$<$20ns for IN11 and
0.01ns$<$t$<$200ns for IN15. %The use of the two different neutron
%spin-echo spectrometers with low and high resolution gives access
%to two different relaxation processes.
Data have been taken at three different temperatures, at
19$^{\circ}$C, in the gel (ripple, P$_{\beta'}$) phase of the
phospholipid bilayers, at 22$^{\circ}$C, just above the
temperature of the main transition in deuterated DMPC-d54 (at
T$_m\approx 21.5$$^{\circ}$C), and at 30$^{\circ}$C, far in the
fluid L$_{\alpha}$ phase of the membranes and above the regime of
so-called anomalous swelling. The corresponding lamellar $d$
spacings were $d$=56\AA, 60\AA\ and 54\AA\ (gel, 22$^{\circ}$C and
fluid), respectively. Two relaxation processes, one at about 10ns
and a second, slower process at about 100ns are clearly
distinguished. Solid lines in the Figure represent least-square
fits of the data to two stretched exponential decays,
$S(q_{||},t)/S(q_{||},0)=(A_1-A_2)\exp[-\left( t/\tau_1
(q_{||})\right)^{\beta1}]+y_2+(A_2-y_2)\exp[-\left( t/\tau_2
(q_{||})\right)^{\beta2}]$. The fitting results for $\beta_{1,2}$
were in the range of $0.96<\beta_{1,2}<1$ for the temperatures
T=22$^{\circ}$C and 30$^{\circ}$C, corresponding essentially to
single exponential relaxations. Contrarily, the gel phase results
(19$^{\circ}$C) give values of $\beta_{1,2}\approx 1.75$, i.e.,
{\em compressed} exponentials. Structural inhomogeneities and
heterogeneous interactions would lead to a local relaxation
dynamics
%that strongly depends on the local structure
and to stretched ($\beta<1$) exponentials. A compressed
exponential decay is incompatible with a diffusive, fluid like
motion of the particles and might therefore be an intrinsic
property of the gel state. The relaxation rates $\tau_1^{-1}$ and
$\tau_2^{-1}$ in the gel and the fluid phase are depicted in
Fig.~\ref{dispersion_fluid_graph.eps} (a) and (b), after
compilation of all measured q$_{||}$ values on the two
spectrometers. Also shown is the fast relaxation process for
T=22$^{\circ}$C.
Both relaxation branches are clearly dispersive. The fast process
shows a q$_{||}^2$ increase at small q$_{||}$ values and a bend at
about q$_{||}\approx$0.015\AA$^{-1}$. The dispersion in the gel
phase and close to the phase transition in
Fig.~\ref{dispersion_fluid_graph.eps} (b) appear to be more
pronounced as compared to 30$^{\circ}$C dispersion. As a
remarkable feature, a soft mode appears in the T=22$^{\circ}$C
dispersion, indicating a significant softening of the bilayer at a
well defined wave number. The slow branches at T=19$^{\circ}$C and
30$^{\circ}$C also show increasing relaxation rates with
increasing q$_{||}$ values, but with a distinct non-polynomial
behavior.

%-------------------------------------------------------------------
% classical smectic Theory and Ribotta fit
The dispersion relation of the fluid L$_{\alpha}$ phase lends
itself to a quantitative comparison with theory. As an effect of
the stacking of the bilayers and the substrate, undulation modes
in lamellar phases decay with relaxation rates
$\tau^{-1}(q_{||})=\kappa/ (\eta_3 d) ~q_{||}^2$, while
fluctuations of a free, non supported, bilayer would decay with
$\tau^{-1} \propto q_{||}^3$ \cite{Zilman:1996,Takeda:1999}.
Generally, {\em undulation} modes are probed at q$_z$ values of
$q_z=2\pi/d$. If the scattering is probed at finite components
$\delta q_z:=(q_z-2\pi/d)$ or measured with a relaxed q$_z$
resolution, {\em baroclinic} modes rather than pure undulations
are probed. Dynamic light scattering experiments can be described
according to bulk smectic elasticity theory and multi lamellar
fluctuations as a function of both symmetry axis $(q_{||},q_z)$ by
the dispersion relation \cite{Freyssingeas:1997}: $\label{sigaud}
\tau^{-1} (q_{||}) = q_{||}^2 ~ \frac{B q_z^2 + (\kappa/d)
q_{||}^4}{\eta_3 q^4+\frac{1}{\mu} q_z^2} ~.$
%Here $B$ denotes the layer compressional modulus (at constant
%chemical potential), and $\eta_3$ the layer sliding viscosity.
The transport coefficient $\mu$ was estimated to $\mu=d^2/(12
\eta_0)$ with $\eta_0$ the viscosity of the the solvent (water)
\cite{Ribotta:1974}. Note that higher order elastic terms have
been neglected. The neutron scattering experiments are carried out
in the first Brillouin zone so that we have to replace q$_z$ by
$\delta q_z$, the deviation from the first zone center. However,
finite size or resolution effects lead to a broadening of the
diffuse Bragg sheet and therefore set a lower cut-off in the
reachable $\delta q_z$. Following the idea of Ribotta
\cite{Ribotta:1974} the above equation is then replaced by
\begin{equation}
\label{ribotta} \tau^{-1}(q_{||}) =\frac{\kappa/d}{\eta_3}
q_{||}^2 \frac{q_{||}^4+(\pi/(\Lambda
D))^2}{q_{||}^4+\frac{1}{\mu\eta_3}(\pi/D)^2} ~.
\end{equation}
In our case, the q$_z$ width of the diffuse scattering is not
defined the finite size $D$ of the lipid film, but comes from the
rather broad instrumental resolution, which is given by the
wavelength spread.
%, and the diffuse scattering is essentially integrated
%in $q_z$, the achievable $q_z$ resolution is not limited by the
%actual finite size $D$ of the film, but by the finite instrumental
%resolution (monochromaticity).
From the measured width $\Delta q_z$ of the first reflectivity
Bragg peak in Fig.~\ref{rocking_fluid_graph.eps} (c) of $\Delta
q_z \simeq 0.015$\AA$^{-1}$ (HWHM, Lorentzian fit), we obtain an
effective finite-size cutoff-length $D=\pi/\Delta q_z\simeq
212$\AA\ that we used to fit the measured fluid dispersion
relation to Eq.~(\ref{ribotta}), see solid line in
Fig.~\ref{dispersion_fluid_graph.eps} (a). The q$_z$ component of
the scattering vector was neglected,
$q=\sqrt{q_z^2+q_{||}^2}\approx q_{||}$, because it may average
out for symmetric broadening. Alternatively, the results can be
compared to the theory given in \cite{Romanov:2002} for thin film
samples, including the boundary effects and surface tension.
However, this theory gives dispersion relations for each of the
$(N-1)$ eigenmodes, which makes a comparison difficult if not
impossible without proper weighting of the modes.

Using Eq.~(\ref{ribotta}), the following results are obtained for
the three free parameters: $\kappa=14.8\pm 8$k$_B$T, $\Lambda=10.3
\pm 2.3$\AA, $\eta_3=0.016\pm 0.0006$Pa~s. B is calculated to
B=1.08~10$^7$J/m$^3$ ($d$=54\AA). Note that the value for
$\Lambda$ compares quite well with the value obtained by a very
different approach but for a similar swelling state
\cite{Ollinger:2005}. Experiments at different osmotic pressures
point to a distinct effect of the swelling state to the
compressional modulus $B$ and $\Lambda$, consequently
\cite{Petrache:1998}. The bending modulus compares quite well to
results from Molecular Dynamics \cite{Lindahl:2000} although the
simulations do not reach the small q$_{||}$ values that we probe
experimentally. $\kappa$ has large errors, but lies in the middle
of the even larger range of literature values \cite{Pabst:2003}.
q$_{||}$ values of q$_{||}>$0.05\AA$^{-1}$ in
Fig.~\ref{dispersion_fluid_graph.eps} (a) and (b) show declining
relaxation rates which deviate from the theoretic curve in
Fig.~\ref{dispersion_fluid_graph.eps} (a). The diffuse Bragg sheet
was slightly bend for the high q$_{||}$ values and the
corresponding diffuse scattering was likely dominated by defect
and no longer purely thermal diffuse scattering with consequences
for the precision in the determination of $\kappa$ and $\eta_3$.
The dynamics at low $q_{||}$ values is governed by the interplay
of viscosity and inertia, i.e., $\tau^{-1}\propto \eta_3/\rho
q_{||}^2$, while at higher q$_{||}$ values, the undulation
dynamics becomes predominant. The difficulty in quantifying
relaxation times at higher q$_{||}$ values thus limits the
precision in the determination of $\kappa$ distinctly. This
constraint can be overcome in future studies with optimized
set-ups and sample preparation. Note that conceptually,
fluctuations in the gel phase, should not be described by the
dispersion relations of fluid smectic A phases. The solid lines in
Fig.~\ref{dispersion_fluid_graph.eps} (b) are therefore guides to
the eye.

%-------------------------------------------------------------
% Discussion and summary
%-------------------------------------------------------------
In summary, NSE provides unique access to collective dynamics of
shape fluctuations in solid supported multilamellar lipid
membranes. %, widely studied as model systems to elucidate structural
%and dynamical properties of their much more complex biological
%counterparts \cite{Lipowsky:1995}.
The dispersion relation of the fast branch with relaxation rates
between 1 and 10ns can be attributed mainly to undulation dynamics
with the expected mixing of baroclinic modes resulting from a 
correction for finite resolution. The resulting effective sliding
viscosity of the membrane system $\eta_3$ was found to be 16 times
higher than that of water. The slow dispersion branch with
relaxation rates of about 100ns may be attributed to a surface
relaxation mode \cite{Bary-Soroker:2006} and will be addressed in
a forthcoming publication. Furthermore a quite localized deviation
from the undulation branch was observed at T=22$^{\circ}$C just
above the main phase transition, indicating the softening of a
well defined wave number of q$_0\approx 0.015$\AA$^{-1}$. From
this result, we speculate that the well known softening of
phospholipid membrane upon approaching the main phase transition
temperature from the fluid phase, i.e., the regime of ''critical
swelling'' or ''anomalous swelling'' \cite{Pabst:2003}, occurs on
a well defined length scale, here at $2\pi/q_{0}\approx 420$\AA.
We have additionally measured the elastic scattering in the ripple
phase and from the satellite peaks of the diffuse Bragg sheet we
determine a ripple periodicity of $d_r\approx$ 130\AA, distinctly
smaller than the length scale of the soft mode observed here.
However, $d_r$ grows significantly at the transition, as we have
measured by atomic force microscopy. % (values of $d_r$=550\AA\
%have been reported for DPPC \cite{Kaasgard:2003}).
Therefore, we speculate that the soft mode in the fluid $L_\alpha$
phase is linked to the formation of the ripple structure in the
$P_{\beta'}$  phase, but this remains to be investigated in more
detail.

%\begin{figure}
%\centering
%\resizebox{0.75\columnwidth}{!}{\rotatebox{0}{\includegraphics{plotsheetgraph.eps}}}
%\caption[]{(Color online). }\label{plotsheetgraph.eps}
%\end{figure}
%\begin{figure}
%\centering
%\resizebox{1.00\columnwidth}{!}{\rotatebox{0}{\includegraphics{ripples_mit_ripplegraph_cdr.eps}}}
%\resizebox{0.5\columnwidth}{!}{\rotatebox{0}{\includegraphics{ripples2.eps}}}
%\caption[]{(Color online). }\label{ripples.eps}
%\end{figure}

%----------------------------------------------------------------------
Acknowledgements: It is a pleasure to thank B.~Farago and
P.~Fouquet for assistance at IN15 and IN11, %T.~Gronemann for help
%with sample preparation,
and the ILL for allocation of ample beam time. T.S. acknowledges
helpful discussions with D.~Constantin. M.C.R enjoyed discussions
with E.~Kats and D.~Bicout.

%------------------------------------------------------------------------------------
%\bibliographystyle{apsrev}
%\bibliography{./Membranes_Rheinstaedter_11052006}

%\bibnamefont{ et~al.}
\end{document}